\documentstyle[12pt]{article}
\def\be{\begin{eqnarray}}
\def\ee{\end{eqnarray}}
\def\ba{\begin{array}}
\def\ea{\end{array}}

\begin{document}

\begin{center}
{\bf\Large {Classical Dynamics of Quantum Numbers with Arrow of
Time}}
\end{center}

\vskip 1cm

\begin{center}
{\bf \large {Vadim V. Asadov$^{\star}$\footnote{asadov@neurok.ru}
and Oleg V.
Kechkin$^{+\,\star}$}\footnote{kechkin@depni.sinp.msu.ru}}
\end{center}

\vskip 2mm

\begin{center}
$^+$Institute of Nuclear Physics,\\
Lomonosov Moscow State University, \\
Vorob'jovy Gory, 119899 Moscow, Russia
\end{center}

\vskip 1mm

\begin{center}
$^\star$Neur\,OK--III\\
Scientific park of MSU, Center for Informational Technologies--104,\\
Vorob'jovy Gory, 119899 Moscow, Russia
\end{center}

\vskip 1cm

\begin{abstract}
We study a quantum theory with complex time parameter and
non-Hermitian Hamiltonian structure. In this theory, the real part
of the complex time is equal to `usual'\, physical time, whereas the
imaginary one is proportional to inverse absolute temperature of the
system. Then, the Hermitian part of the Hamiltonian coincides with
conventional operator of energy; the anti-Hermitian part, which is
taken as a symmetry operator, defines decay parameters of the
theory. We integrate the equations of motion in a Hamiltonian proper
basis, and detect a classical dynamics of the corresponding quantum
numbers, using their continuous approximation and the zero limit of
the Plank's constant. It is proved, that this dynamics possesses a
well defined arrow of time in the isothermal and adiabatic regimes
of the thermodynamical evolution of the system.
\end{abstract}

\vskip 0.5cm

PACS No(s).\, : 05.30.-d,\,\,05.70.-a.

\renewcommand{\theequation}{\thesection.\arabic{equation}}

\vskip 1cm

\section{Introduction}
\setcounter{equation}{0}

Irreversibility (in time) is one of the most evident and mysterious
property of the real world. One says about `arrow of time'\, in
evolution of the Universe and its closed subsystems to stress this
fundamental non-equivalence between the real past and future
\cite{atf}--\cite{atl}, and operates by different kinetic and
thermodynamical constructions to describe the corresponding
unidirectional processes \cite{td}. The main problem in this
activity is related to realization of the formal thermodynamics on
the base of some fundamental dynamical theory, like the standard
quantum or classical mechanics. Note, that this theory must be taken
without any `hand made'\, arrow of time -- i.e., it must be
conservative in the conventional sense of the canonic formalism.
Thus, one needs to derive, for example, the irreversible
thermodynamics from the definitely reversible mechanics, which seems
non-realistic programm in the framework of any consistent
mathematics. Of course, one can try to improve this hopeless
situation by corresponding modification of the fundamental theory
itself. Note, that it is exactly the way that we have chosen in our
approach to solve the problem under consideration.

In \cite{q-1}, we have developed the modified quantum theory, which
possesses arrow of time in various regimes of its dynamics. In fact,
this theory provides a general theoretical framework for unification
of the standard quantum and statistical mechanics. Its dynamics is
described in terms of the complex evolutionary parameter
$\tau\neq\tau^*$, and the non-Hermitian Hamiltonian operator
$\cal{H}\neq\cal{H}^+$. For the natural analogies of the standard
conservative systems, we put ${\cal{H}}_{,\tau}=0$. The main
dynamical equation of the theory is postulated in the its
conventional Schr\"{o}dinger's form, \be\label{G3} i\hbar
\Psi_{,\tau}={\cal{H}}\Psi. \ee  We consider the holomorphic variant
of the theory, with $\Psi_{,\tau^{*}}=0$ and
${\cal{H}}_{,\tau^{*}}=0$, which leads to the simplest
generalization of the standard theoretical quantum scheme. Also, we
restrict it by the theories with $\left [
\cal{H},\,\,\cal{H}^+\right ]=0$. The last relation becomes an
identity in the standard quantum theory case, when
${\cal{H}}={\cal{H}}^+$, so our approach can be understood as the
`minimal generalization'\, of the standard one.

The complex evolutionary parameter $\tau$ can be parameterized in
terms of the real variables $t$ and $\beta$ in the following form:
\be\label{G5} \tau=t-\,i\,\frac{\hbar}{2}\,\beta, \ee whereas the
non-Hermitian operator ${\cal{H}}$ can be represented using the
Hermitian operators $E$ and $\Gamma$ as \be\label{G6}
{\cal{H}}=E-\,i\,\frac{\hbar}{2}\,\Gamma. \ee Note, that
\be\label{G4'}[E,\,\,\Gamma]=0, \ee in accordance to the restriction
imposed above. In \cite{q-1} it was argued, that, if one identifies
the quantities $t$ and $E$ with the `usual'\, time and energy
operator, respectively, then the remaining quantities $\beta$ and
$\Gamma$ mean the inverse absolute temperature $\beta=1/kT$
(multiplied to the Bolzman constant $k$), and the operator of
inverse decay time parameters of the system. Then, this scheme of
generalization of the quantum theory must be completed by
introducing of a conception of thermodynamic regime, which has a
form of fixation of the temperature function $\beta=\beta (t)$. The
simplest regimes are the isothermal $\beta=\rm{const}$ and the
adiabatic $\bar{E}(t,\beta)=\rm{const}$ ones (we use bar for all
averaged quantities here). In \cite{q-1} it was shown, that for
these two regimes the function $\bar \Gamma=\bar \Gamma(t,
\beta(t))\equiv \bar \Gamma[t]$ is non-increasing. This fact
indicates the presence of arrow of time in the corresponding
evolution of the system. In \cite{q-2} it was studied the quantum
dynamics with the operator $\Gamma$ of the parity type. For this
theory, the left-right asymmetry becomes the result of its
unidirectional dynamics. The Lyapunov function $\Gamma[t]$ detects
it in the explicit form for the arbitrary thermodynamic regime.

In this work, we check an existence of the irreversible classical
dynamics of the fundamental type, which can be derived from the
quantum theory described above. We use for its formulation and study
the most convenient set of the original quantum variables -- the set
of quantum numbers, related to the corresponding stationary
Schr\"{o}dinger's problem. We consider this set in the natural
continuous approximation, and perform the limit procedure
$\hbar\rightarrow 0$ in the appropriate form for the all theory
structures. We define classical dynamics as the dynamics of maximum
of the resulting probability density of the system. In doing so, we
derive equations of motion (of the Hamilton's type), which describe
the dynamics of this maximum point, and prove the irreversible
character of this dynamics. Namely, we show, that in the isothermal
and adiabatic regimes this modified classical dynamics leads to the
non-increasing classical function $\Gamma[t]$, which must be
calculated on the classical trajectory under consideration.

\section{Hamilton Equations for Quantum Numbers}

\setcounter{equation}{0}

Let us express all significant quantities of the theory in terms of
a common basis of the eigenvectors $\psi\left (n\right)$ of the
commuting operators $E$ and $\Gamma$. We take it in the orthonormal
form (i.e., we mean, that the identity $\psi\left
(n\right)^+\psi(k)=\delta(n,\,k)$ takes place). Here, of course, the
indexes are understood in the appropriate `multi-sense'\, (and all
summations are of the corresponding type). The eigenvalue problem
under consideration reads: \be\label{S1} E\psi\left (n\right)=E\left
(n\right)\psi\left (n\right), \qquad \Gamma\psi\left
(n\right)=\Gamma\left (n\right)\psi\left (n\right); \ee it can be
reformulated in terms of the non-Hermitian operator ${\cal H}$.
Actually, it is easy to see, that $\psi\left (n\right)$ is the
eigenvector for this operator, which corresponds to the complex
eigenvalue ${ H}\left (n\right)=E\left
(n\right)-i\hbar/2\,\Gamma\left (n\right)$. Then, the state vectors
$\Psi\left (\tau,\,n\right)=\exp[-i{ H}\left
(n\right)\tau/\hbar]\psi\left (n\right)$ satisfy the
Schr\"{o}dinger's equation (\ref{G3}), and also form the complete
(but $\tau$-dependent) basis. This basis can be used for
representation of any solution $\Psi(\tau)$ of the Schr\"{o}dinger's
equation in the form of linear combination with some set of constant
parameters $C\left (n\right)$, i.e., as $\Psi\left (\tau\right
)=\sum_n C\left (n\right)\Psi\left (\tau,\,n\right)$ Using this
decomposition formula and the orthonormal basis properties, one can
calculate the probability $P\left (t,\,\beta,\,n\right)$ to find the
quantum system in its basis state $\Psi\left (\tau,\,n\right )$,
when it is described by the state vector $\Psi(\tau)$. The result
reads: \be\label{S2} P\left (t,\,\beta,\,n\right)=\frac{w\left
(t,\,\beta,\,n\right)}{Z}, \ee where $Z=Z(t,\,\beta)=\sum_n w\left
(t,\,\beta,\,n\right)$, and  \be w(t,\beta,\, n)&=&\exp{\left [-{
S_2(t,\beta,\, n)}\right ]}, \nonumber\\{ S_2(t,\beta,\, n)}&=&
\sigma(n)+E\left (n\right)\beta +\Gamma\left (n\right) t.\ee Here we
have put $\sigma(n)=-\log |C(n)|^2$.

Now let us put, for convenience, $\tilde t_1=t,\, \tilde t_2=\beta$,
and also $\tilde E_1=\Gamma,\, \tilde E_2 =E$, and combine the
corresponding quantities to the sets $\tilde t_\alpha$ and $\tilde
E_\alpha$ ($\alpha=1,2$). We define the classical value
$n_c=n_c(\tilde t_\alpha)$ of the collective $n$-variable according
to the relation \be P\left ( \tilde t_\alpha,\,\, n_c\right )=
\max_{n}P\left ( \tilde t_\alpha,\, n\right ).\ee Thus, we identify
the classical theory with the theory which describes the dynamics of
the `maximum point'\, of probability density. In the `hard'\,
classical limit the probability density is described by the
delta-functional distribution. It is clear, that the resonance
character of the classical probability density must be guaranteed by
the corresponding choice of the weight parameter $\sigma\left
(n\right)$. Actually, this parameter is some initial data of the
theory, so it can be taken in the appropriate form at the beginning
of evolution of the system.

We derive the defining $n_c$-relation using differentiation of the
necessary extremum condition $\dot{ S}_2\left [\tilde t,\, n(\tilde
t)\right ]=0$ in respect to $\tilde t_{\alpha}$: \be\label{diff}
\frac{\partial}{\partial_{\tilde t_{\alpha}}}\dot{ S}_2\left [\tilde
t,\, n(\tilde t)\right ]=0\ee (here the dot means differentiation in
respect to the $n$-components). Note, that this $\tilde t_{\alpha}$
- differentiation is understood with taking into account the total
$\tilde t_{\alpha}$-dependence of the corresponding quantity. It
fact, Eq. (\ref{diff}) means a conservation of the extremal
character of the classical trajectory of the system during its
possible physical evolution.

It is not difficult to prove, that the dynamical equation for the
quantity $n_c$, which follows from Eq. (\ref{diff}), reads:
\be\label{n} n_{,\,\tilde t_{\alpha}}=-{ A}_2^{-1}\dot {\tilde
E}_{\alpha},\ee where ${ A}_2=\ddot{
S}_{2\,c}=\ddot{\sigma}_{c}+\sum_{\beta}\ddot{\tilde
E}_{\beta\,\,c}\tilde t_{\beta }$. Here $\dot {\tilde E}_{\alpha}$
is the column of the derivatives of the first order  $\partial
\tilde E_{\alpha}/{\partial { n_{k}}}$, whereas the quantity (for
example) $\ddot{ S}_{2\,c}$ means the symmetric matrix which is
constructed from the derivatives of the second order $\partial {
S}_{2\,c}/{\partial { n_{k}}\partial { n_{l}}}$.

Then, the limit $\hbar\rightarrow 0$ is important for calculation of
the set of canonical moments, which we define as \be p=\dot{{
S}_1}\left [\tilde t_{\alpha},\, n(\tilde t_{\alpha})\right ].\ee In
this formula, ${ S}_1$ means the real part of the phase ${ S}$ of
the wave-function $\Psi$ in the $n$-representation, which is given
by the relation $\Psi \left (\tilde t_{\alpha},\, n\right )=C\left
(n\right)\Psi(\tau,\, n)$. It is not difficult to prove, that \be {
S}_1\left (\tilde t_{\alpha},\, n\right )=\lambda\left
(n\right)-E\left (n\right)t+\frac{\hbar^2}{4}\Gamma\left
(n\right)\beta.\ee Here the function $\lambda \left (n\right)$ is
defined according to the relation $ C\left (n\right)=| C\left
(n\right)|\exp{\left (\frac{i}{\hbar}\lambda \right )}$. Then, in
the explicit form, and after the taking of the limit mentioned
above, one concludes, that \be p=\dot{\lambda}-\dot{E} t.\ee Then,
for the time derivatives of the moment variable $p$ it is possible
to establish the equation $p_{,\,\tilde t_{\alpha}}=-\dot
{E}\delta_{1,\alpha}+{ A}_1n_{{,\tilde t}_{\alpha}}$, where ${
A}_1=\ddot{ S}_{1\, c}=\ddot{\lambda}-\ddot{E}\,t$. Finally, using
the relation (\ref{n}), one concludes, that \be\label{p}
p_{,\,\tilde t_{\alpha}}=-\dot {E}\delta_{1,\alpha}-{ A}_1{
A}_2^{-1}\dot{\tilde E}_{\alpha}.\ee It is clear, that the relations
(\ref{n}) and (\ref{p}) form the pair of Hamilton's equations for
the theory under consideration. We state, that this system is
consistent, i.e., that the mixed time derivatives for the coordinate
and moment variables are equal to the inverse ones. This means, that
this system of classical equations is correct.

\section{Arrow of Time in Dynamics of Quantum Numbers}

\setcounter{equation}{0}

Our main statement is related to irreversibility of the evolution of
the system, defined by the Hamilton's equations (\ref{n}),
(\ref{p}), and by the temperature regime $\beta=\beta(t)$ taken in
the appropriate form. Below we consider the isothermal and adiabatic
cases, as it had been performed for the original quantum theory in
\cite{q-1}. To prove the presence of arrow of time in these
physically important regimes, let us choose the representation,
where $\Gamma=\Gamma(n_1)$ and $E=E(n_2)$. It is clear, that this
special representation always exists for the theory taken (i.e., for
the starting quantum theory with the commuting energy and decay
operators). From the system of Hamilton's equations it follows, that
\be\label{syst} \ba {ccc} n_{1, \,t}=-\Gamma^{'}({ A}_2^{-1})_{11},
&\quad& n_{2,\, t}=-\Gamma^{'}({
A}_2^{-1})_{21},\\n_{1,\,\beta}=-E^{'}({
A}_2^{-1})_{12},&\quad&n_{2, \,\beta}=-E^{'}({
A}_2^{-1})_{22},\ea\ee where the prime means derivative in respect
to the corresponding (single) variable of the function under
consideration. Then, for the total $t$-derivative of the function
$\Gamma$, i.e., for the quantity $d\Gamma/dt=\Gamma^{'}d
n/dt=\Gamma^{'}(n_{1,\,t}+\beta^{'} \, n_{1,\,\beta})$, one can
calculate its explicit form using Eq. (\ref{syst}). In the
isothermal case (with $d\beta/dt=0$), the result reads: \be
\frac{d\Gamma}{dt}=-\left (\Gamma^{'}\right )^2\left ({
A}_2^{-1}\right )_{11}.\ee In the adiabatic regime (with $dE/dt=0$),
one obtains, that \be \beta^{'}=-\frac{\Gamma^{'}}{E^{'}}\frac{\left
({ A}_2^{-1}\right )_{21}}{\left ({ A}_2^{-1}\right )_{22}}\ee for
its `temperature curve',\, and that
\be\frac{d\Gamma}{dt}=-\frac{\left (\Gamma\right )^2}{\left ({
A}_2^{-1}\right )_{22}}\left |\ba{ccc}\left ({ A}_2^{-1}\right
)_{11}&\,&\left ({ A}_2^{-1}\right )_{12}\\\left ({ A}_2^{-1}\right
)_{21}&\,&\left ({ A}_2^{-1}\right )_{22}\ea\right |\ee for the
total time derivative of the decay function $\Gamma$. It is clear,
that $d\Gamma/dt<0$ in the both situations considered, because the
coefficient $\left ({ A}_2^{-1}\right )_{11}$ and the determinant
written above are positive quantities in view of the positive
definiteness supposed for the quantity $d^2 { S}_2$ on the whole
classical trajectory. Thus, one deals with the strictly decreasing
evolution of the decay function $\Gamma[t]$ for the arbitrary
initial data taken. This proves the Lyapunov nature of this
function, and the presence of arrow of time in the dynamics of the
classical system under investigation.

\section{Conclusion}

In this work we have detected an existence of the classical dynamics
of the Hamilton's type, which possesses the well-defined arrow of
time (in the isothermal and adiabatic regimes of the thermodynamical
evolution, at least). In fact, it is established the unified theory
of classical and statistical mechanics, which gives a natural
realization for the standard classical thermodynamics. We have used
for its derivation the original set of quantum numbers, which
originate from the general solution of the stationary
Schr\"{o}dinger's equation. Our starting quantum theory is taken as
holomorphic in respect to the complex parameter of evolution; the
dynamics is governed by the non-Hermitian Hamiltonian structure. We
would like to note, that the results obtained in this work are the
most general ones for the all theories under consideration.
Actually, one can choose the `quantum numbers representation'\, for
the arbitrary theory of this type.

In the next publication we will develop the classical dynamics with
arrow of time in terms of the arbitrary set of canonical variables
of the theory. We think, that this modified canonical formalism can
give the base for construction of fundamental and time-irreversible
quantum field and string theories in the simple and natural way.
Also, it seems actually promising for applications in particle
physics, gravity and cosmology \cite{for}-\cite{PartPhysCosm-f}.

\vskip 10mm \noindent {\large \bf Acknowledgements}

\vskip 3mm \noindent We would like to thank prof. B.S. Ishkhanov for
many discussions and private talks which were really useful for us
during this work preparation. One of the authors (O.V.K) was
supported by grant ${\rm MD \,\, 3623.\, 2006.\, 2}$.

\end{document}